\documentclass[prb,preprint,aps,longbibliography,groupedaddress]{revtex4-1}

\usepackage{bbm}
\usepackage{graphicx}
\usepackage{subfigure}
\usepackage{amsmath}
\usepackage{booktabs}
\usepackage{ulem}
\usepackage{xcolor}

\begin{document}

\title{Advantage of utilizing nonlocal magic resource in Haar random circuits}

\author{Xiao Huang$^{1}$, Guanhua Chen$^{1}$, and Yao Yao$^{1,2}$\footnote{Electronic address:~\url{yaoyao2016@scut.edu.cn}}}

\address{$^1$ Department of Physics, South China University of Technology, Guangzhou 510640, China\\
$^2$ State Key Laboratory of Luminescent Materials and Devices, South China University of Technology, Guangzhou 510640, China}

\date{\today}

\begin{abstract}
Magic resources and entanglement are fundamental components for achieving the universal quantum computation, so is the interplay between them. Herein, we uncover an intrinsic scaling law of the magic resource and bond dimension of matrix product states in Haar-random quantum circuits, that is, the magic resource is converged on a bond dimension in logarithmic scale with the system size. From a practical perspective, this finding substantially enhances the classical simulability of nonstabilizerness. It also allows us to utilize the bond dimension as a bridge to link the entanglement and the nonlocal magic resource, which extends the capacity perspective that the entanglement plays the role of container for the nonlocal magic resource. Furthermore, the intrinsic scaling enables an information separation between the nonlocal magic resource and the extra entanglement. This, in turn, leads to the conclusion that, any dynamical relation between magic and entanglement resources is ruled out. In other words, it is inappropriate to regard the entanglement as the driving force of the growth and spreading of nonlocal magic resource.
\end{abstract}

\maketitle

\section{\label{sec:intro}INTRODUCTION}

The resources from quantum entanglement and magic states promise to us with the universal quantum computation \cite{resource}, by enabling the solution of computational tasks with potential exponential speedups over classical computers, which is well known as quantum supremacy or advantage \cite{shor1,quantum_supremacy}. Although stabilizer circuits have been able to attain highly entangled states, the celebrated Gottesman-Knill theorem tells us that, the circuits solely with Clifford operations can be efficiently simulated on a classical computer \cite{Gottesman1,Gottesman2}. The magic resource is therefore proposed to realize non-Clifford gates, under the architecture of stabilizer error corrections \cite{kitaev1}. The nonstabilizerness, a measure of magic resource, then quantifies the amount of non-Clifford resources to reach the target state, which typically requires complicated distillation protocols \cite{kitaev1,distillation1,distillation2}. Subsequently it is significant to investigate how magic resources build up and manifest potential facilitation on near-term quantum devices.

The stabilizer R\'{e}nyi entropy (SRE) \cite{SRE,magicMPS}, and similarly the mana entropy \cite{mana,manaEntropy}, serve as the effective measures for quantifying magic resources. However, it is conventionally difficult for these measures to be utilized on scalable quantum systems, due to the exponential scaling number of Pauli strings. Great effort has thus been devoted to the method of scalable measures of magic resource. For example, Lami and Collura introduced a perfect sampling method to efficiently quantify nonstabilizerness of matrix product state (MPS) at a computational complexity of $O(N\chi^{3})$ with $N$ being the system size and $\chi$ the bond dimension of MPS \cite{Pauli_sampling}. This benchmark inspires us to further comprehend the classical computational circumstances for magic state dynamics in different quantum circuits, such as measurement-induced phase transitions \cite{monitor1,monitor2} and magic resource spreading \cite{Turkeshi_2025}. However, in systems where entanglement grows rapidly, the efficiency of this sampling algorithm becomes severely limited, as the bond dimension scales exponentially with the entanglement entropy, leading to a computational cost that is effectively exponential in the entanglement. Our main results show that, when the object is solely constrained to extract magic-related quantities rather than the complete information of each quantum state, the computational cost can be dramatically reduced, even in regimes with intrinsically high entanglement.

As minimal models for chaotic dynamics, Haar-random brick-wall circuits are a benchmarking playground to study the universal dynamics of quantum resources \cite{RQC,RQC1,RQC2,RQC3,RQC4,scrambling}, which exhibit rich quantum complexity stemming from quantum randomness. A common approach to generate Haar-random states points to the Haar-random brick-wall circuits, in which magic resources have been demonstrated to saturate on a timescale that is in logarithmic relation with the system size, distinct from the entanglement that saturates on a timescale linear in the system size\cite{Turkeshi_2025,RQC3,ent_increase}. Notably, single-qubit Haar-random gates can also generate magic resource, but the resulting nonstabilizerness remains strictly local because of the absence of nonlocal gates. Meanwhile, one can notice that the magic resource in product Haar-random states is always less than that in the saturated $N$-qubit Haar-random states. By comparison, it is the two-qubit Haar-random gates that generate both the entanglement and nonlocal magic resource, delocalizing and driving the system to saturated $N$-qubit Haar-random states.

A number of recent studies focused on the interplay between magic resource and entanglement, and the entanglement is conventionally regarded as playing an important role in the growth and spreading of magic resources, especially the nonlocal nonstabilizerness. Hou \textit{et al.} demonstrated that the entanglement can act as a highway that facilitates the spreading of locally injected magic resource, by designing an analytically tractable setup which separates the process of magic injection from entanglement generation \cite{highway}. Turkeshi \textit{et al.} demonstrated that under ergodic many-body dynamics, magic resources equilibrate on timescales logarithmic in the system size \cite{Turkeshi_2025}, in sharp contrast to entanglement, which grows ballistically and thus much more slowly. This raises an important question: does the growth of entanglement drive the growth of nonlocal nonstabilizerness toward its saturation value? As shown below, the central result of this work provides a clear and compelling resolution to this question.

Very recently, Li \textit{et al.} introduced a so-called \textit{water-pouring} analogy, in which the entanglement plays the role of a container and the magic resource acts as its content \cite{watercup}. This remarkably insightful idea was formulated within the setting of measurement-based quantum computation and graph states, and our work attempts to extend this conceptual innovation to generic many-body states and chaotic circuits. Frau \textit{et al.} studied the magic resource in MPS with limited bond dimensions for spin-1 anisotropic Heisenberg chains \cite{bond}, and found that nonstabilizerness converges more easily than entanglement in practical computations. This result motivates us to investigate how entanglement accommodates magic within the MPS framework, since the bond dimension serves as a discrete yet tunable parameter linking entanglement to nonlocal nonstabilizerness. This, in turn, enables a clear quantitative analysis of how entanglement holds the magic resource, ultimately leading to the entanglement nonstabilizerness capacity that captures this intricate relationship.

The paper is structured as follows. In Sec. \ref{sec:1}, the definitions of the stabilizer R\'{e}nyi entropy (SRE), the nonlocal nonstabilizerness and von Neumann entropy are formulated. In Sec. \ref{sec:2}, we introduce Haar-random brick-wall circuits and numerical simulation methods. In Sec. \ref{sec:3}, we introduce two numerical experiments and design a tractable setup separating the process of magic resource injection from generation of extra entanglement which does not contribute to the magic resource growth and spreading. In Sec. \ref{sec:level1}, we present our results on the long-time limit of SRE, which reveals an intrinsic scaling between magic resource and bond dimension in Haar-random circuits. Finally in Sec. \ref{Conclusions}, we summarize this work.

\section{\label{sec:measure}Methodology}
\subsection{\label{sec:1}Formula of stabilizer entropy and entanglement}
Let us consider an $N$-qubit chain. A Pauli string is identified as $\sigma = \prod_{j=1}^{N} \sigma_{j} \in \mathcal{P}_{N}$, and $\mathcal{P}_{N} = \{\sigma^{0}, \sigma^{1}, \sigma^{2}, \sigma^{3}\}^{\otimes N}$ is the set of $N$-qubit Pauli operator strings. For a pure normalized state $\rho$, a widely used measure of nonstabilizerness is the stabilizer $n$-R\'{e}nyi entropy (SRE) defined as
\begin{equation}
{M}_{n} = \frac{1}{1-n} \ln \sum_{\sigma \in \mathcal{P}_{N}} \frac{1}{2^{N}} \text{Tr}[\rho \sigma]^{2n}.
\label{eq:SRE}
\end{equation}
It satisfies the following properties. (i) Faithfulness: ${M}_{n}$ vanishes for stabilizer states and is positive for other states. (ii) Stability under Clifford unitary operations $U_{\rm C}$: ${M}_{n}(U_{\rm C}\rho U_{\rm C}^{\dagger}) = {M}_{n}(\rho)$. (iii) Additivity: ${M}_{n}(\rho \otimes \rho') = {M}_{n}(\rho) + {M}_{n}(\rho')$. It is worth noting that SRE can be explained by the inverse participation ratio, which also results in the participation entropy. Significantly, both SRE and the participation entropy exhibit remarkably similar evolutionary behavior while approaching Haar-random states in random unitary circuits. This similarity suggests that SRE can be a characterization of Hilbert space delocalization \cite{delocalization}.

Furthermore, it is noticed that the R\'{e}nyi entropy at rank $1$ comprises a non-negative real-valued function $\Xi = (1/2^{N}) {\rm Tr}[\rho \sigma]^{2}$, which can be safely interpreted as a probability distribution over all Pauli strings, as it sums to unity. This enables an alternative idea to replace the iterative calculation of $4^{N}$ Pauli operator strings to the statistical sampling. Lami and Collura recently introduced this method to evaluate the SRE via perfect Pauli sampling with MPS techniques \cite{Pauli_sampling}, in which the variance has an upper bound in the order of $O(1/\mathcal{N})$ for $n > 1$, with $\mathcal{N}$ being the sampling number. It has also been shown that, this method can be efficiently applied to the R\'{e}nyi entropy at any rank.

Due to the strong connection between entanglement and nonlocal magic resource, it is important to clarify the precise meaning of the latter, in order to distinguish the local and nonlocal components of nonstabilizerness. By nonlocal magic resource, we refer to the component of the SRE that fundamentally depends on quantum correlations (e.g., entanglement) and therefore cannot be localized on individual subsystems. In other words, it is the part of the SRE that vanishes when all nonlocal correlations are removed.

Based upon the nonlocal feature of nonstabilizerness, we define it for the Haar random states as follows:
\begin{equation}
{M}_{\mathrm{NL}}\bigl(|\psi_{\mathrm{Haar}}^{(N)}\rangle\bigr)
= {M}\bigl(|\psi_{\mathrm{Haar}}^{(N)}\rangle\bigr)
- {M}\bigl( \bigotimes_{i=1}^{N} |\phi_{\mathrm{Haar}}^{(1)}\rangle \bigr),
\label{eq:nonlocal_SRE}
\end{equation}
where the first term on the right-hand side represents the SRE of the $N$-qubit Haar random states, and the second term corresponds to the SRE of the direct product of $N$ single-qubit Haar random states. We notice that the second term on the right-hand side depends only on the system size $N$, while the first term depends on both $N$ and the bond dimension $\chi$. This indicates that when analyzing the relation between bond dimension and nonlocal nonstabilizerness, the second term acts merely as a baseline and can be omitted in the fitting analysis. Therefore, we can safely use the first term to capture the feature of nonlocal nonstabilizerness resulting from changes in bond dimension $\chi$.

We concurrently use the von Neumann entropy as a measure of entanglement $S$, which is given by,
\begin{equation}
S = -\operatorname{Tr} \left[ \rho_{l} \log_{2} \left( \rho_{l} \right) \right],
\end{equation}
where $l \in \{1, \cdots, n-1\}$ denotes the cutting position of the subsystems, and $\rho_{l}$ is the relevant reduced density matrix by partially tracing out the sites from $l+1$ to $n$, namely $\rho_{l}={\rm Tr}_{[(l+1)\cdots n]}[\rho]$. $S(\rho)$ usually takes its maximum value cutting at around the center of the chain, so we select this maximum value at each step in the whole calculations \cite{Efficient}.

\subsection{\label{sec:2}Haar-random circuits and simulation}

Haar-random brick-wall circuits are a benchmarking playground for quantum many-body physics and a tractable setting to explore universal collective phenomena far away from equilibrium. This universality directly arises from the combination of high randomness and locality, ensuring that the dynamics are independent of the specific details of the system. Based on this model, numerous studies have uncovered the features of quantum thermalization and revealed universal dynamics of quantum information \cite{RQC,RQC1,RQC2,RQC3,RQC4,scrambling}. As minimal models for chaotic dynamics, the phenomenology exhibited by Haar-random circuits can be typically extended to a broad class of chaotic many-body systems.

\begin{figure}[htbp]
    \centering
    \includegraphics[scale=0.6]{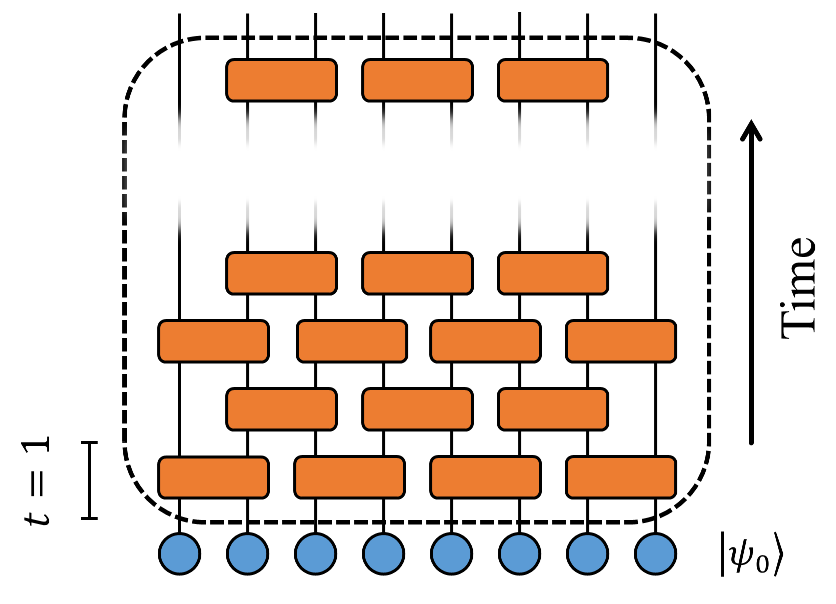}
    \caption{Structure of Haar-random brick-wall circuits, with only two-qubit gates. The initial state is product state. Each layer contributes a depth (time step) of unity.}
    \label{fig:circuit}
\end{figure}

We mainly study the relationship between the long-time limit of nonstabilizerness and bond dimension in Haar-random brick-wall circuits with open boundary conditions, as sketched in Fig.~\ref{fig:circuit}. As the basic structure, local two-qubit Haar-random gates generate both the entanglement and nonstabilizerness, moving the initial product state to Haar-random states. A unit depth comprises one layer of two-qubit Haar-random gates applied to odd or even bonds by turns, and two-qubit Haar-random gates are chosen independently with the Haar distribution on the unitary group $\mathcal{U}(d^{2})$, where $d$ denotes the dimension of local Hilbert space. Each depth of circuits corresponds to a unit discrete time step.

In practical quantum experiments, the two-qubit Haar-random gate is usually replaced by a structure that comprises two random single-qubit rotations in parallel and a randomly directed CX \cite{Koh_2023}. The nonstabilizerness evolution of these two circuits is highly similar, but we have to point out that due to the specific structure of the experimental setup, it is easy to find that the local nonstabilizerness reaches full capacity at the very first step, and the remaining evolution is in fact governed by the growth and spreading of nonlocal magic resource. By analogy, one may reasonably expect that the overall magic evolution of Haar-random brick-wall circuits essentially depends on the nonlocal nonstabilizerness increase, while the contribution from local nonstabilizerness merely serves as a baseline, which reinforces our definition of nonlocal nonstabilizerness in Sec. \ref{sec:1}.

We simulate the random circuits composed of two-qubit Haar-random gates by MPS and evaluate the numerical results via Pauli-based sampling method \cite{TEBD1,TEBD2}. The bond dimension of MPS is controlled in two modes. The infinite mode refers to that the bond dimension is not limited, namely the upper bound is infinite, and the finite mode means there is a preset finite upper bound. Below the upper bound the required bond dimension is automatically adjusted by the precision of entanglement, and in our calculations, we set the precision of singular value decomposition to be $10^{-8}$.

The latter provides a practical approach to calculate the SRE of MPS based on perfect Pauli sampling, and for the parameter sets we consider herein, it is found that ${M}_{n}$ generally attains a stable value within fewer than $\mathcal{N}=3000$ samples. Considering an $N$-qubit state described by MPS with finite bond dimension $\chi$, the Pauli-based sampling method has a computational complexity scaling as $\mathcal{O}(\mathcal{N} N\chi^{3})$, while the computation of von Neumann entropy via MPS has a complexity scaling as $\mathcal{O}(d^{3}N\chi^{3})$. For the qubit in our case $d=2$, so that for $\mathcal{N} \gg d^{3}$ the computational cost of the nonstabilizerness is substantially higher than that of entanglement.

We mainly focus on the long-time limit of nonstabilizerness and entanglement, averaged over a large number of quantum random circuit realizations. It is necessary to distinguish the quenched and the annealed averages. The former is the average characterization over the ensemble of trajectories, while the latter is the characterization of the average state. However, the quenched and the annealed averages of nonstabilizerness are always approaching to each other with increasing number of qubits $N$ and time steps $t$, specifically for $N = 8$ qubits \cite{Turkeshi_2025}. Moreover, the quenched averages are more suitable for Pauli sampling approach, which are then chosen in our calculations.

\subsection{\label{sec:3}Setup of numerical experiments}

Recent studies have shown that obtaining converged results for nonstabilizerness is typically easier than entanglement \cite{bond}. Based on the close relationship between the entanglement and bond dimension, researchers investigated the nonstabilizerness with limited bond dimension for the ground state of spin-1 anisotropic Heisenberg chains. Concretely, three different transitions are considered: Haldane-N\'eel, Haldane-large D, and large D-XY \cite{XXZ}. The convergence of SRE for all three distinct transitions is observed at small bond dimensions, while the entanglement entropy requires a larger bond dimension to converge. Furthermore, the SRE density $m$ exhibits a linear dependence on $1/\chi^{2}$ as
\begin{equation}
m(1/\chi^{2})=m_{0}+m_1/\chi^{2},
\end{equation}
where $m_0$ and $m_1$ are fitting parameters. Similarly in Haar-random circuits evolution, SRE saturates exponentially in time, while the entanglement exhibits ballistic increase \cite{RQC3,ent_increase}. This different convergence rate of magic resource and entanglement stems from their different physical pictures. The SRE denotes the global properties of a state, whereas the entanglement entropy accounts for the correlated information between subsystems.

Although several works have attempted to establish a direct relation between the two quantum resources \cite{flatness,Hamma}, the approaches they employ are often intricate and difficult to be further developed. Remarkably, studying how the entanglement holds the magic resource within the MPS framework provides a decisive advantage: the bond dimension serves as a natural bridge linking entanglement to nonlocal nonstabilizerness, enabling a clear quantitative analysis. A very recent work noticed that relatively small bond dimension could underestimate the true value of SRE \cite{msm2-vmg7}, a fact that is little known. In the following, we quantify how small a bond dimension is sufficient in Haar-random circuits; this turns out to reflect a universal convergence rate between nonstabilizerness and bond dimension. Such insight into the rapid convergence of the SRE offers substantial practical benefits for classical simulation, which is directly relevant to our topic. The significance of our work even goes far beyond this practical aspect. Li \textit{et al.} recently introduced the idea that entanglement serves as a container for magic within graph states \cite{watercup}, and our work extends this conceptual innovation to generic many-body states and chaotic circuits.

We design two types of numerical experiments as follows.

\textit{Experiment 1.-} In this experiment, we consider a one-dimensional chain of $N$ qubits under Haar-random circuits. We control the bond dimension $\chi$ in the finite mode during the Haar-random circuits simulated up to a depth of 40, which is sufficiently deep to produce Haar random output states. For various system sizes $N$ (up to $100$ qubits) and bond dimensions $\chi$ (ranging from $1$ to $30$), we simulate Haar-random circuit evolution and evaluate the SRE ${M}_{1}$ and ${M}_{2}$ of output states utilizing Pauli-based sampling method \cite{Pauli_sampling}. When the system size $N\leq 20$, ${M}_{n}$ is evaluated with $\mathcal{N}=10^{4}$ samples; when $N>20$, $\mathcal{N}=3\times 10^{3}$ samples are used. We compute the mean values $\bar{M}_{1}$ and $\bar{M}_{2}$ by averaging over 500 quantum circuit realizations for each $N$ and $\chi$ when the system size $N\leq 60$. We observe that the magic resource converges at an extraordinarily small bond dimension $\chi_{\text{SRE}}$ compared to that for entanglement, as detailed below.

\textit{Experiment 2.-} The easier convergence of SRE uncovered by \textit{Experiment 1} enables separating the process of magic resource injection from the extra entanglement which does not contribute to the magic resource increase. As small bond dimension $\chi_{\text{SRE}}$ suffices to accurately capture the SRE of Haar-random states, the truncation of bond dimension at $\chi_{\text{SRE}}$ could solely influence the entanglement that is reserved in the discarded states. The retained entanglement could potentially play a role on the magic resource increase, and we can justify the accuracy by comparing the coefficients of the saturation rates with recent works \cite{Turkeshi_2025}.

In this experiment, we simulate Haar-random circuits with several system sizes $N=11,15,30,40,50$, with the corresponding finite bond dimension $\chi_{\text{SRE}}=15,16,17,20,23$ selected from the calculations in \textit{Experiment 1}. The simulations up to a depth of 20 are based on the MPS, and meanwhile we compute ${M}_{1}$ and ${M}_{2}$ during evolution utilizing Pauli-based sampling method. The quenched average nonstabilizerness $\bar{M}_{1}$ and $\bar{M}_{2}$ are obtained by averaging over $500$ trajectories of random circuits realizations for each $N$. Additionally, we simulate a system of 11 qubits with full bond dimension $\chi=2^{N/2}=32$ to ensure the MPS simulation is precise. The consistent evolution of magic state with various bond dimensions justifies the reliability of our protocol.

\section{\label{sec:level1} Results}
\subsection{\label{sec:a} Convergence rate of magic resource and entanglement}

\begin{figure}[htbp]
    \centering
    \includegraphics[scale=0.8]{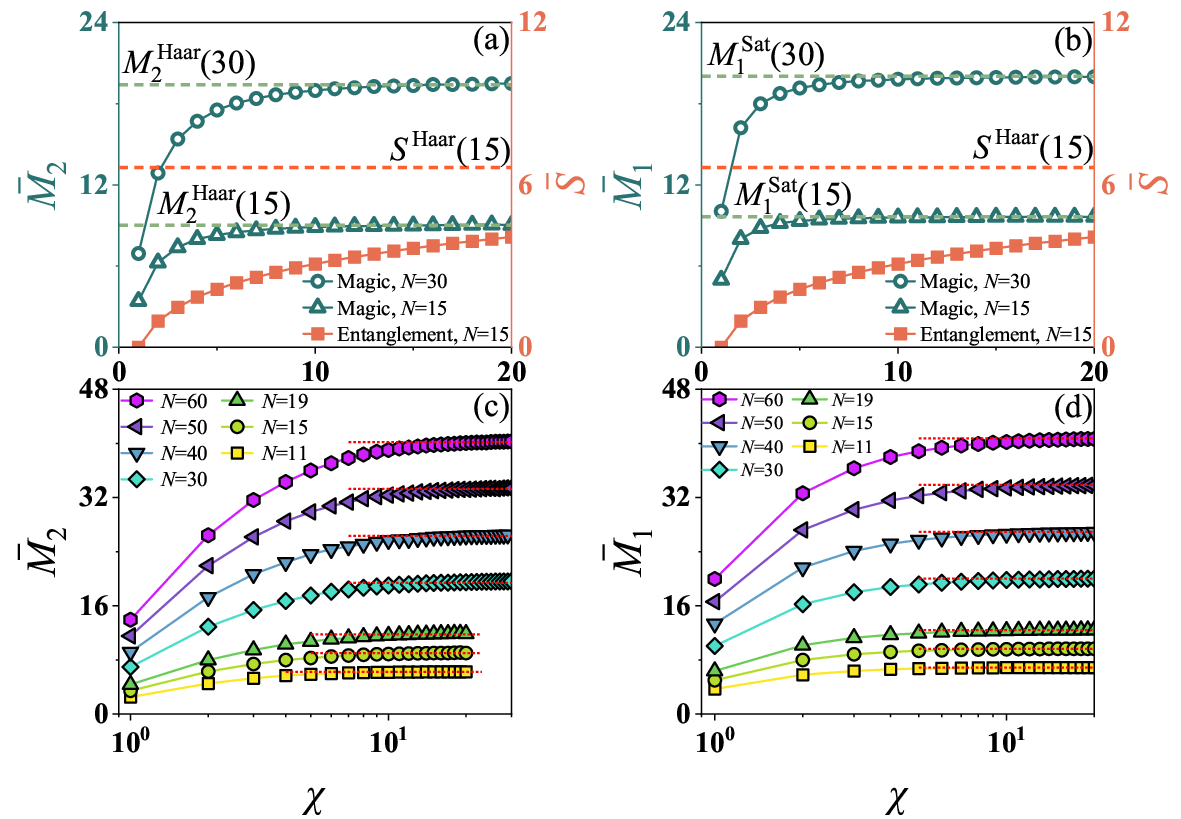}
    \caption{
        The averaged SRE $\bar{M}_{n}(N)$ of Haar-random states as a function of the system size $N$ for the variable bond dimension $\chi$.
        (a)(b) The different convergence curves of SRE $\bar{M}_{n}$ and entanglement $\bar{S}$.
        (c)(d) The SRE $\bar{M}_{n}$ with more system sizes, all of which universally exhibit sharp increase to ${M}_{n}^{\mathrm{Sat}}$.
        The dashed lines represent the SRE of $N$-qubit Haar-random states. For $n=2$, ${M}_{2}^{\mathrm{Sat}}={M}_{2}^{\mathrm{Haar}}$. For $n=1$, the maximum value of $\bar{M}_{1}$ is selected as the approaching convergence value ${M}_{1}^{\mathrm{Sat}}$.
    }
    \label{fig:bondSRE}
\end{figure}

As shown in Fig.~\ref{fig:bondSRE}, we find that the averaged SRE converges at small bond dimensions for various system sizes, while the entanglement continues increasing. For ${M}_{2}$, one can derive an explicit form for $N$-qubit Haar-random states as given by
\cite{satM2}
\begin{equation}
{M}_{2}^{\mathrm{Haar}} \equiv -\ln \left[ \frac{4}{2^{N} + 3} \right],
\label{eq:M_Haar}
\end{equation}
which is labeled in the figure by dashed lines respectively, corresponding to relevant convergence values of $\bar{M}_{2}$ for $N$-qubit system. For ${M}_{1}$, however, there is not an explicit form, so instead we use the maximum value of $\bar{M}_{1}$ denoted by ${M}_{1}^{\mathrm{Sat}}(N)$, which has been close enough to the convergence value. In Fig.~\ref{fig:bondSRE}(a), it is observed that, a bond dimension of $ \chi\sim 11 $ is sufficient for an accurate computation of $\bar{M}_{2}$ for system size $N=15$, and for $N=30$, $ \chi\sim 14 $ is enough. Due to the bond dimension constraint, however, $\bar{S}$ for $N=15$ and $N=30$ almost overlap, so we draw the entanglement for $N=15$ lonely, which is still far from convergence. According to the formula of von Neumann entropy, $\bar{S}$ is expected to converge at a bond dimension $\chi\sim2^{N/2}$, which is much larger than that of $\bar{M}_{2}$. In Fig.~\ref{fig:bondSRE}(b), an even faster convergence curve of $\bar{M}_{1}$ is also observed. Both $\bar{M}_{1}$ and $\bar{M}_{2}$ follow the same path of convergence, and the different convergence rates can be reflected by the fitting parameters as discussed soon. The variation of convergence rates for $M_{n}$ stems from their information distributing on the bond dimension. Compared to ${M}_{2}$, the information of ${M}_{1}$ is more concentrated within a smaller bond dimension. In Fig.~\ref{fig:bondSRE}(c) and (d), we show that for systems of up to $60$ qubits, $\bar{M}_{1}$ and $\bar{M}_{2}$ reach convergence by a bond dimension of around $\chi=20$. This indicates for both ${M}_{1}$ and ${M}_{2}$, and particularly for larger system sizes, there is a significant difference in bond dimension requirements for evaluating nonstabilizerness and entanglement.

\begin{figure}[htbp]
	\centering
    \includegraphics[scale=0.8]{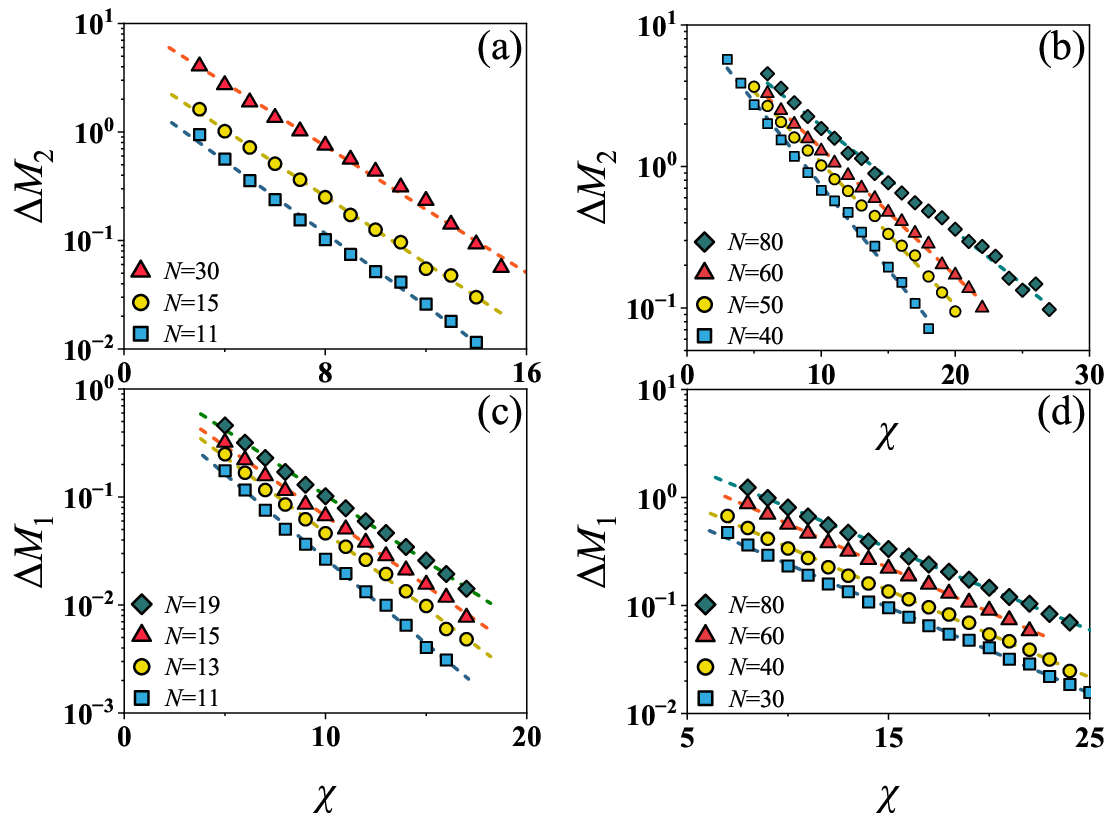}
	\caption{Scaling of the SRE deviation $\Delta {M}_{n}$ with the bond dimension $\chi$. The different panels correspond to various R\'{e}nyi rank $n$ and different system sizes $N$:
	(a) $n=2,N=11,15,30$,
	(b) $n=2,N=40,50,60,80$,
	(c) $n=1,N=11,13,15,19$,
	(d) $n=1,N=30,40,60,80$.
	The dashed lines represent fittings.}
	\label{fig:logbondSRE}
\end{figure}

In order to further establish the formula of the convergence rate, we define $\Delta {M}_{n}(\chi)$ as the deviation between the $\bar{M}_{n}(\chi)$ of Haar-random states with a given bond dimension $\chi$ and the convergence value ${M}_{n}^{\mathrm{Sat}}$, which is expressed as
\begin{equation}
\Delta {M}_{n}(\chi)=| {M}_{n}^{\mathrm{Sat}} - \bar{M}_{n}(\chi)|.
\label{eq:Delta_M}
\end{equation}
As shown in Fig. \ref{fig:logbondSRE}, it is clear that $\ln(\Delta {M}_{n})$ exhibits a pretty universal linear dependence on $\chi$ for all calculated system sizes, which turns out to be the essential result of this work. The dashed lines represent fitting of the form $\ln(\Delta {M}_{n}) = -{\alpha}_{n}\chi + {\beta'}_{n}$, with $\alpha$ and $\beta'$ the fitting parameters, indicating an exponentially declining relation between $\Delta {M}_{n}$ and $\chi$,
\begin{equation}
\Delta {M}_{n} = {\beta}_{n} \exp(-{\alpha}_{n}\chi),
\label{eq:expx1}
\end{equation}
where ${\beta}_{n}=\exp({\beta}_{n}')$. It is noted that ${M}_{2}^{\text{Haar}}$ is not explicitly the upper bound of $\bar{M}_{2}$ for Haar-random states, so $\bar{M}_{2}$ could slightly fluctuate around ${M}_{2}^{\text{Haar}}$. More samplings can reduce the fluctuation. In terms of Eq.~(\ref{eq:expx1}), the exponential decline in bond dimension $\chi$ means the rapid buildup of $\bar{M}_{n}$ in the initial stage of evolution. Notably, as the saturation value, for $n=2$, ${M}_{2}^{\text{Sat}}$ approaches $N\ln(2)$ as $N\rightarrow\infty$. A similar volume law exists for $n=1$. When $\chi=1$, the system is a product Haar-random state, with $\Delta {M}_{n}$ being proportional to the system size $N$, following a volume law. When $\chi>1$, $\bar{M}_{n}$ approaches the volume scaling of ${M}_{n}^{\text{Sat}}$. Since these two terms are proportional to the system size $N$, the difference $\Delta {M}_{n}$ likewise follows a volume law.

\begin{figure}[htbp]
	\centering
	\includegraphics[scale=0.8]{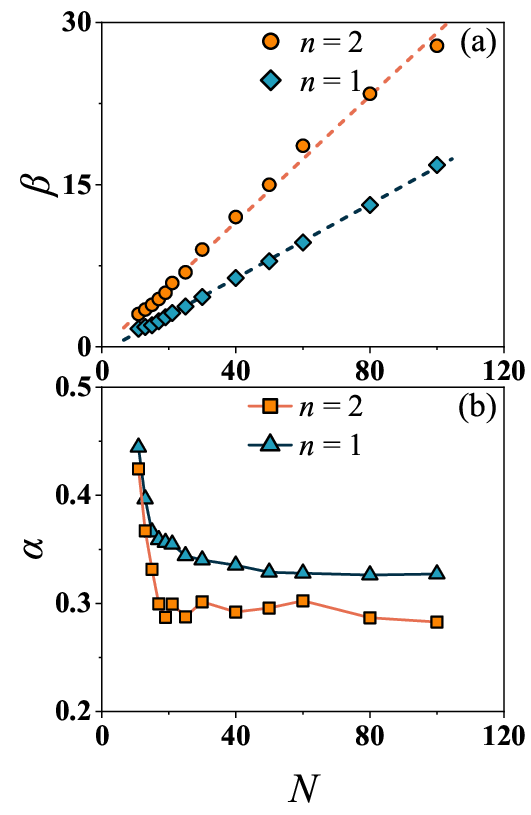}
	\caption{
		Fitting parameters in the exponential decline of SRE deviation with various system sizes $N$.
		(a) $\beta$ increases linearly with the system size $N$. For $n=2$, the slope is ${\lambda}_{2}=0.29$, and for $n=1$, ${\lambda}_{1}=0.17$.
		(b) As $N$ increasing, $\alpha$ remains  at around ${\alpha}_{2}\approx0.29$ for $n=2$ and ${\alpha}_{1}\approx0.33$ for $n=1$.
	}
	\label{fig:betaN}
\end{figure}

Fig.~\ref{fig:betaN}(a) shows the linear dependence of ${\beta}_{n}$ on the system size $N$ up to 100 qubits, i.e., ${\beta}_{n}(N)={\lambda}_{n} N+{\mu}_{n}$, where ${\lambda}_{n}$ and ${\mu}_{n}$ are fitting parameters and the latter can be ignored in the large $N$ limit. From information-theoretic considerations, the linear scaling of ${\beta}_{n}$ with $N$ is expected, as ${M}_{n}$ behaves as an extensive quantity. This linear dependence indeed reflects that the nonstabilizerness generated by Haar-random circuits grows simultaneously across all qubits. Thus the SRE deviation is expressed as,
\begin{equation}
\Delta {M}_{n}={\lambda}_{n} N\exp(-{\alpha}_{n}\chi).
\label{eq:expx}
\end{equation}
In Fig.~\ref{fig:betaN}(b), it is observed that ${\alpha}_{1}$ consistently exceeds ${\alpha}_{2}$, a trend that agrees with the faster convergence rate of $\bar{M}_{1}$ previously noted in Fig.~\ref{fig:bondSRE}.

As a consequence, for sufficiently large bond dimension, the deviation $\Delta {M}_{n}$ is proportional to the system size $N$ and declines exponentially with $\chi$. Alternatively speaking, the convergence bond dimension scales logarithmically with system size $N$, serving as the essential conclusion of the present work. It implies that the magic resource is converged on a bond dimension that is in logarithmic scale with the system size, completely different from the exponential scaling for entanglement.

From the perspective that entanglement serves as the container for nonlocal magic resource, this result leads to the following heuristic insights. When the bond dimension $\chi$ of an MPS grows linearly, the entanglement entropy merely increases logarithmically, as $\bar{S}=\text{log}_{2}({\chi})+{c}$, this follows directly from combining Haar randomness with the finite subspace dimension $\chi$, and is easily verified numerically; $c$ is a constant here. Meanwhile, nonlocal nonstabilizerness approaches its stauration value exponentially fast in $\chi$, as
\begin{equation}
\bar{M}_{n}(\chi)
= {M}_{n}^{\mathrm{Sat}} - \lambda_{n} N\exp(-{\alpha}_{n}\chi)
= {N}\!\left( {c}_{n} - \lambda_{n} \exp(-\alpha_{n}\chi) \right),
\end{equation}
${c}_{n}$ is a constant for large $N$. Thus, we could define the capacity of entanglement to store nonstabilizerness,
\begin{equation}
{C}=\frac{d \bar{M}_n}{d \bar{S}}
= N \lambda_n \alpha_n \cdot \chi \exp(-\alpha_n \chi).
\label{eq:capacity}
\end{equation}
In the regime where the bond dimension $\chi$ increases from $1$ up to $\chi_\text{SRE}$, the nonstabilizerness is extensive, whereas the entanglement can be safely regarded as an intensive quantity. Because $\partial_\chi {C}({N},\chi)=(1-\alpha_n \chi)\exp(-\alpha_n \chi)=0$, the optimal bond dimension ${\chi}^*=1/\alpha_n$ and the maximum specific capacity ${C}^*= \lambda_n$, which reflects the special structure of Haar-random circuits.

The linear dependence of capacity $C$ on $N$ originates from the fact that the SRE captures global properties of the state. This behavior indicates that the nonlocal nonstabilizerness generated by Haar-random circuits grows in the meantime across all qubits, resulting in a uniform distribution of nonstabilizerness in Haar-random states. It is worth noting that, for Haar-random states in a one-dimensional chain, the entanglement structure is not uniform: the entanglement entropy is larger near the center and smaller near the edges. The fact that such a non-uniform entanglement structure can nevertheless support a uniformly distributed nonlocal nonstabilizerness is itself nontrivial.

From the perspective of simulation power, the bond dimension $\chi$ required heavily determines the cost of classical computations in the MPS framework. As well known, the family of tensor computation has become the currently most powerful numerical methods in the study of one-dimension many-body system \cite{DMRG1,DMRG2,TEBD1,TEBD2}. However, to fully capture the high entanglement, a bond dimension that scales exponentially with the system size is required, leading classical MPS simulations to hit the entanglement exponential wall.

In contrast, our results indicate that the classical simulability of magic resources can be substantially stronger, because when one targets nonstabilizerness, a bond dimension that scales only logarithmically with the system size is sufficient. Under this favorable scaling, the computational complexities of TEBD and the Pauli-based sampling method become ${O}({N} \text{ln}^3 (N))$, i.e., polynomial in $N$. Notably, we observe that the sampling number $\mathcal{N}$ required for the convergence of SRE estimation does not increase once the system size becomes sufficiently large $(N\ge30)$. This behavior originates from the concentration of the Haar measure, under which the fluctuations of the SRE with circuit realizations are strongly suppressed as $N$ grows. Consequently, utilizing nonlocal nonstabilizerness as a characterization provides clear advantages for random circuit simulation particularly in regimes with inherently high entanglement.

\subsection{\label{sec:b} Random unitary circuits with limited bond dimensions}

Exploring various setups to achieve separating the process of magic resource injection from the entanglement generation, has been a central theme in recent studies on the interplay between these two resources \cite{highway,TCC,TCC1,monitor1,monitor2}. For example, Hou \textit{et al.} have shown that the stabilizer entanglement facilitates the spread of locally injected magic resource \cite{highway}. Here now we consider an alternative setup, which can be called the information separation between the process of magic resource injection and the generation of extra entanglement. Based on the magic resource convergence bond dimension $\chi_{\text{SRE}}$, we establish the dynamics of SRE in the long-time limit of Haar-random circuits merely using a small bond dimension. Truncating the bond dimension at a small value diminishes most of entanglement that can be called extra entanglement, while the retained entanglement may still play a role.

To address the question raised in the beginning, we first have to clarify what we mean by ``drive" or ``driving force". We say the nonlocal nonstabilizerness is driven by the entanglement when there exists an explicit functional dependence between them under ergodic many-body dynamics, of the form $M(t)=f(S(t))$. Moreover, the results of Turkeshi \textit{et al.} do not rule out this possibility during the early stages of circuit evolution \cite{Turkeshi_2025}.

\begin{figure}[htbp]
	\centering
	\includegraphics[scale=0.8]{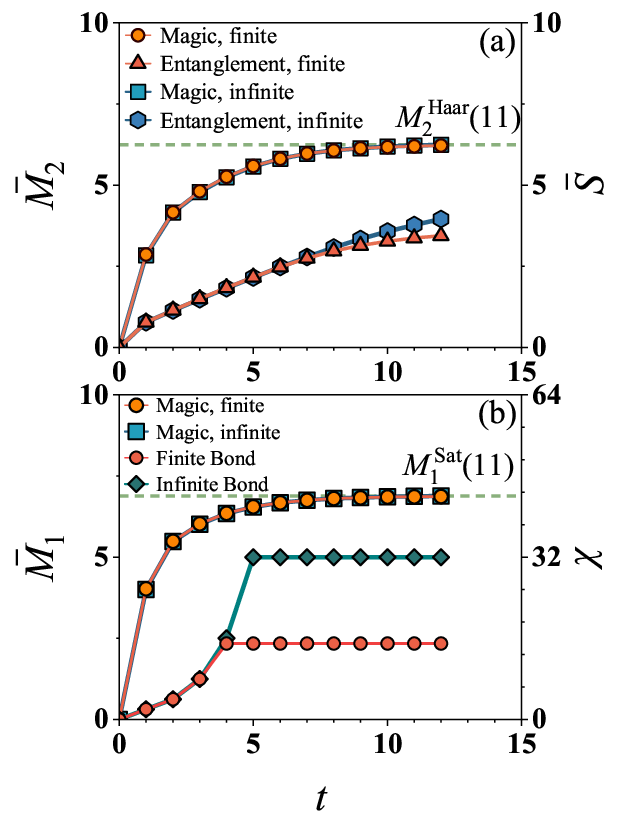}
	\caption{
		Evolution of Haar-random circuits on an 11-qubit system with various bond dimensions. Wherein, ``finite" stands for finite bond dimension, and ``infinite" stands for infinite bond dimension.
		(a) Bond dimension for magic resource convergence $\chi_{\text{SRE}}=15$ can faithfully establish the dynamics of $\bar{M}_{2}$, but fail to completely establish entanglement.
		(b) The dynamics of $\bar{M}_{1}$ and the change of required bond dimensions under random circuits are displayed.
	}
	\label{fig:compare11}
\end{figure}

As shown in Fig.~\ref{fig:compare11}, the evolution of $\bar{M}_{n}$ is almost identical for different bond dimensions. In Fig.~\ref{fig:compare11}(a), $\bar{M}_{2}$ of both cases concurrently saturates at $t\approx10$. In contrast, the curves for entanglement entropy start to diverge at around $t=8$. In Fig.~\ref{fig:compare11}(b), the dynamics of $\bar{M}_{1}$ also exhibits similar results. Moreover, the required bond dimension below the preset upper bound is also targeted, whose growth diverges at time $t=4$. Before reaching the preset bound, the required bond dimension increases exponentially with time.

\begin{figure}[htbp]
	\centering
	\includegraphics[scale=0.8]{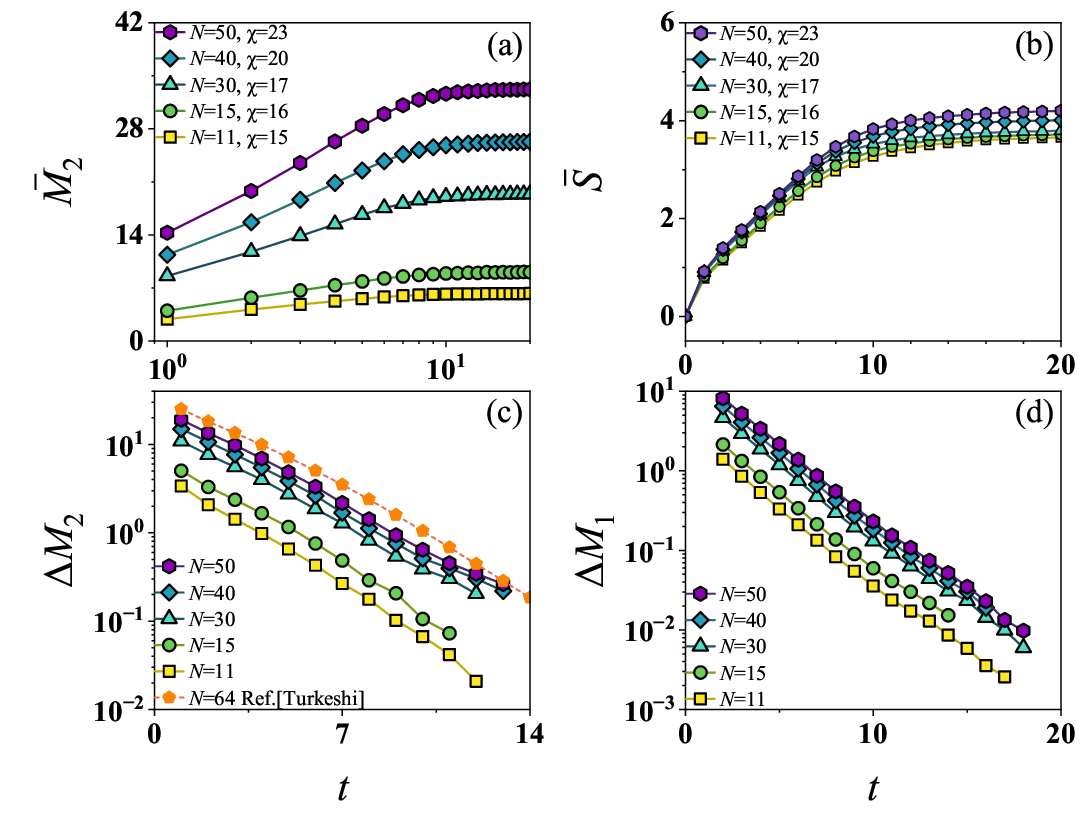}
	\caption{
		The SRE and entanglement evolution in Haar-random circuits for various system sizes $N$ with corresponding limited bond dimensions $\chi_{\text{SRE}}$.
		(a) The averaged SRE $\bar{M}_{2}$ sharply converges to ${M}_{2}^{\mathrm{Sat}}$.
		(b) $\bar{S}$ exhibits a ballistic increase, with saturation time depending on $\chi_{\text{SRE}}$.
		(c)(d) The SRE deviation exhibits exponential declines $\Delta {M}_{n} \propto N e^{-\gamma_{n} t}$ with $\gamma_{n}$ being a fitting parameter.
	}
	\label{fig:evolution}
\end{figure}

We further study the faithfulness of our protocol for larger system sizes. As shown in Fig.~\ref{fig:evolution}(a) and (c), $\bar{M}_{2}$ sharply rises up and saturates to $M_{2}^{\mathrm{Haar}}$, while the deviation $\Delta {M}_{2}$ declines exponentially with time as $\Delta {M}_{2} \propto N e^{-\gamma_{2} t}$. When the system size $N \geq 30$, the fitting parameter $\bar{\gamma_{2}} \approx 0.43$, which coincides with that in the exponential relaxation of ${M}_{2}$ obtained by Turkeshi \textit{et al.} \cite{Turkeshi_2025}. To enable a more direct comparison, we further include the publicly available dataset at $N=64$ from Turkeshi \textit{et al.}, which is plotted as the orange curve in Fig.~\ref{fig:evolution}(c). This indicates that the small bond dimension merely suffices to bear with the dynamics of ${M}_{2}$ in Haar-random circuits. In Fig.~\ref{fig:evolution}(d), $\Delta {M}_{1}$ exhibits similar exponential decline to their stable values under the evolution of random unitary circuits. As a result, the SRE deviation $\Delta {M}_{n}$ is proportional to the system size $N$ and declines exponentially with time $t$, which is expressed as
\begin{equation}
\Delta {M}_{n} \propto N \exp(-{\gamma}_{n} t).
\label{eq:expt}
\end{equation}
Alternatively speaking, the SRE saturates at a time logarithmic with system size $N$. We subsequently prove that the dynamics of magic resource is faithfully captured by MPS with an extraordinarily small bond dimension, as $\chi_{\text{SRE}}\propto \ln(N)$. Considering the similarity between Eq.~(\ref{eq:expx}) and (\ref{eq:expt}), we can safely conclude that
\begin{equation}
\chi_{\text{SRE}} \sim t_{\text{SRE}}.
\label{eq:simi}
\end{equation}
This same scaling of bond dimension and time is expected, because both of them originate from the same underlying mechanism. Namely in Haar-random circuits, each qubit contributes to the nonstabilizerness in the meantime, regardless of whether the growth is governed by increasing $\chi$ or by increasing $t$.

Let us then investigate the co-evolution of magic resource and entanglement with the threshold bond dimension $\chi_{\text{SRE}}$. It is clear that the dynamics of magic state is independent of the component of entanglement captured by the bond dimension beyond $\chi_{\text{SRE}}$, and $\bar{M}_{n}$ saturates at time $t_{\mathrm{SRE}}\propto \ln(N)$. Fig.~\ref{fig:evolution}(b) displays the dynamics of averaged entanglement $\bar{S}$ with the bond dimension constraint. Due to the ballistic increase of entanglement entropy in ergodic many-body dynamics \cite{RQC3,ent_increase}, $\bar{S}$ saturates at time $t_{\mathrm{ENT}}\propto \log_2(\chi_{\text{SRE}})$. Given that $\chi_{\text{SRE}}\propto \ln(N)$, we thus derive a double logarithmic scaling of entanglement with time as
\begin{equation}
t_{\mathrm{ENT}}\propto \log_2(\ln(N)).
\label{eq:tent}
\end{equation}
The bunching shape of saturation values in Fig.~\ref{fig:evolution}(b), in contrast to the antibunching shape in Fig.~\ref{fig:evolution}(a), further manifests these distinct scalings. In addition, when the system size $N \rightarrow \infty$, $t_{\mathrm{ENT}} \ll t_{\mathrm{SRE}}$. Remarkably, as mentioned in Sec. \ref{sec:1}, the overall magic evolution of Haar random circuits essentially depends on the increase of nonlocal nonstabilizerness. That is, for sufficiently large systems, the entanglement reaches saturation much earlier than nonlocal magic resource, implying the absence of any dynamical relation between these two quantum resources. In other words, it is inappropriate to regard the entanglement as the driving force of the growth and spreading of nonlocal magic resource.

Together with the fact that entanglement acts as the container for nonlocal nonstabilizerness, we see that the double-logarithmic scaling of the entanglement saturation time given by Eq.~(\ref{eq:tent}) quantifies the temporal duration required for the system to build up sufficient entanglement to hold the nonlocal magic resource. As Haar-random circuits are minimal models for chaotic dynamics, we point out that our findings, both the logarithmic scaling of magic resource with bond dimension, and the causality between nonlocal magic resource and entanglement, are believed to be the universal phenomena in a wide class of chaotic many-body systems.

\section{\label{Conclusions} Conclusions}

In summary, focusing on the long-time limit of Haar-random circuits, we have investigated the relationship between the magic resource and the bond dimension. For the system size being up to $N=100$, we observe the magic resource converging on a bond dimension that is in logarithmic scale with the system size, as $\chi_{\text{SRE}}\propto \ln(N)$. On one hand, our results reveal that the classical simulability of nonstabilizerness can be substantially enhanced, since the computational complexity remains polynomial in the system size even in regimes with intrinsically high entanglement. On the other hand, we systematically investigate how entanglement acts as a container for nonlocal nonstabilizerness under ergodic many-body dynamics, and introduce the concept of specific capacity to capture this intricate relationship.

Based on this intrinsic scaling, we design a tractable setup achieving an information separation between the process of magic resource injection and the generation of extra entanglement. Remarkably, in random circuits evolution with bond dimension constraint, the nonlocal magic resource saturates on a timescale that is in logarithmic relation with the system size, distinct from the entanglement that saturates on a timescale doubly logarithmic in the system size. These intrinsic scaling laws indicate the absence of dynamical relation between entanglement and nonlocal nonstabilizerness.

\begin{acknowledgments}
The authors gratefully acknowledge support from the National Natural Science Foundation of China (Grant No.~12374107).
\end{acknowledgments}

\bibliography{magic_v16.bbl}

\end{document}